\documentstyle[11pt,newpasp,twoside,epsf]{article}

\begin{document}

\title{Dark Matter in Dwarf Galaxies: The First Dark Galaxy?}
\author{Joshua D. Simon, Timothy Robishaw, and Leo Blitz}
\affil{Department of Astronomy, University of California at Berkeley, 
601 Campbell Hall, CA  94720}

\begin{abstract}
We present new H {\sc i} observations of the high-velocity cloud (HVC)
that we resolved near the Local Group dwarf galaxy LGS~3.  The cloud
is rotating, with an implied mass that makes it dark matter-dominated
no matter what its distance from the Milky Way is.  Our new,
high-sensitivity Arecibo observations demonstrate that the faint H
{\sc i} features that we previously described as tidal tails are
indeed real and do connect to the main body of the HVC.  Thus, these
observations are consistent with our original hypothesis of a tidal
interaction between the HVC and LGS~3.  We suggest that the HVC may be
one of the missing dark matter satellites in the Local Group that are
seen in Cold Dark Matter numerical simulations but have not yet been
identified observationally.
\end{abstract}

\vspace{-0.1in}
\section{Introduction}

The first hint of unusual goings-on in the neighborhood of LGS~3 came
from Christian \& Tully (1983), who mentioned that H {\sc i}
observations by Hulsbosch revealed that LGS~3 lies on the edge of a
large cloud of gas (called HVC 127-41-330 by them) that contains a
significant velocity gradient.  However, with the 36\arcmin\
resolution of his Dwingeloo observations, Hulsbosch was apparently
unable to draw any firm conclusions about the nature of this object.
Over the next 17 years the situation remained murky as few observers
paid attention to this part of the sky.  The sole reference to the
cloud during this period was in the HVC catalog of Hulsbosch \& Wakker
(1988), who assumed that it was part of LGS~3.

Three years ago the cloud was rediscovered by Blitz \& Robishaw (2000)
during their search for gas associated with Local Group dwarf
spheroidals in the Leiden/Dwingeloo Survey (LDS) of Galactic Neutral
Hydrogen (Hartmann \& Burton 1997).  Blitz \& Robishaw (2000) pointed
out the offsets in position and velocity between this cloud and the H
{\sc i} known to be directly associated with LGS~3 (Thuan \& Martin
1979; Young \& Lo 1997).  They speculated that the cloud might have
been ram-pressure stripped out of LGS~3 by hot gas in the halo of M31.
They noted, however, that the velocity of the cloud should be less
negative than the velocity of LGS~3 ($v_{\mbox{{\tiny HVC}}} > -287$
km~s$^{-1}$), which it is not ($v_{\mbox{{\tiny HVC}}} \approx -330$
km~s$^{-1}$).  Like Hulsbosch, their ability to consider other
possible origins for the cloud was limited by the low angular
resolution of the LDS data.

Robishaw, Simon, \& Blitz (2002) shed new light on this object with a
high-resolution, wide-field H {\sc i} map of the region made at
Arecibo.  These observations showed that the cloud is completely
separate from LGS~3, demonstrating that it is indeed an HVC.

\section{Discussion}

The Robishaw et al. (2002) observations raised several pressing
questions.  What is the nature of the LGS~3 HVC?  Are LGS~3 and the
HVC actually located at the same distance, or is their apparent
proximity just a chance alignment?  Even though there is no H {\sc i}
connection between LGS~3 and the HVC, could the high-velocity gas be
associated with LGS~3 in any way?

\subsection{Distance of the HVC}

In Robishaw et al. (2002) we pointed out that the probability of a
chance superposition between a compact HVC and a dwarf galaxy --- if
they are completely unrelated objects --- is rather low.  Using the
statistics of HVCs in the LDS (de Heij et al. 2002) and known Local
Group dwarfs, we showed that the likelihood of an HVC being located
less than 30\arcmin\ from a dwarf is $\sim4\%$ in this area of the
sky.  Given such an angular coincidence, the probability of the
velocities also matching within 50 km~s$^{-1}$ is $\sim20\%$, for a
joint probability of less than $1\%$.  This line of reasoning
suggests, but does not prove, that the HVC and LGS~3 are physically
associated.

A more powerful argument can be made based on the faint H {\sc i}
strips that our first Arecibo map revealed on either side of the HVC.
These features appeared to be long, thin tails of gas that connect to
the HVC.  That appearance, along with their nearly symmetrical
placement relative to the line between LGS~3 and the HVC, is a strong
signal of a tidal interaction between the two objects.  If this
interpretation is correct, the HVC must be located physically close to
LGS~3, at a distance of about 700 kpc from the Milky Way.  This HVC
would then be the first HVC known to lie more than $\sim50$ kpc away.

\subsection{HVC Rotation Curve}

The HVC shows a surprisingly regular velocity gradient indicative of
circular rotation.  Robishaw et al. (2002) argued against the
possibility of a chance alignment of two separate clouds moving at
slightly different velocities, so if the gradient is \emph{not} due to
rotation, the only alternative is shear.  A simple timing calculation
shows that this explanation is unlikely.  The cloud currently has a
radius of about 3 kpc (D/1 Mpc), where D is the distance of the HVC.
The velocity gradient is 15 km~s$^{-1}$, which would correspond to an
expansion velocity of 7 km~s$^{-1}$.  At that expansion rate, the time
for the cloud to reach its current size is $4 \times 10^{8}$ (D/1 Mpc)
yr.  In this scenario, distances of less than 100 kpc are highly
implausible --- the HVC would have to be extraordinarily young, and
even then the nearest possible progenitors, the Magellanic Clouds, are
too far away for the HVC to have reached its present position at any
reasonable velocity.  At 700 kpc the lifetime of the cloud is still
uncomfortably shorter than a Hubble time, and the lack of a credible
progenitor object is equally severe.  We conclude that the velocity
gradient is almost certainly due to rotation.

\begin{figure}[!t]
\plotone{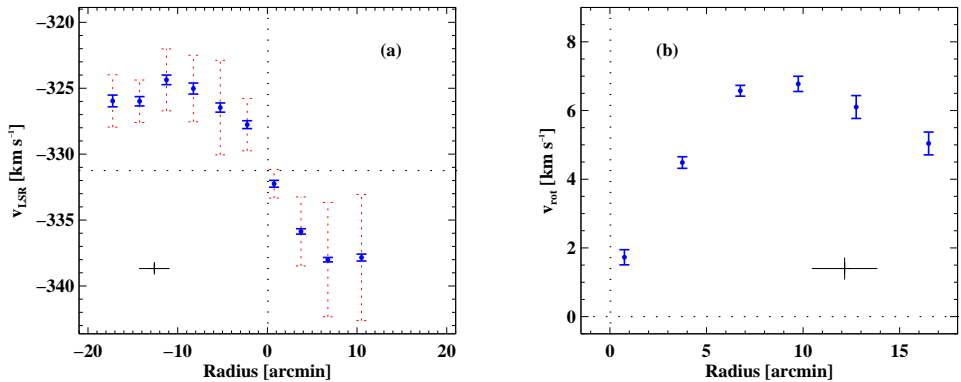}
\caption{(a) Major axis rotation curve of the HVC.  The blue (red)
error bars represent the uncertainties in the Gaussian fits (the
scatter in the velocity field perpendicular to the major axis).
Removing this rotation curve from the velocity field yields a residual
field with an rms of 3 km~s$^{-1}$.  (b) The rotation curve folded
about the point of maximum symmetry.  After a solid-body rise out to
10\arcmin, the rotation curve turns over and begins to decline.  The
error bars are the 1~$\sigma$ uncertainties of the weighted average of
the Gaussian fits.  The crosses indicate the velocity and spatial
resolution of our measurements.
\label{lgs3rc}}
\end{figure}

The rotation curve of the HVC, as measured by Robishaw et al. (2002),
is symmetric and flattens out at a radius of about 10\arcmin\ (see
Fig. \ref{lgs3rc}).  The total mass derived from the rotation curve is
larger than the H {\sc i} mass for any reasonable distance, implying
that the HVC is dark matter-dominated.  At 700 kpc, 82 \% of the mass
is in dark matter; at 50 kpc, 99 \% of the mass is dark.  The HVC can
only be dynamically dominated by luminous material if its distance is
$\ga 2$ Mpc, which is extremely unlikely.

\subsection{New Observations}

\begin{figure}[!t]
\plotone{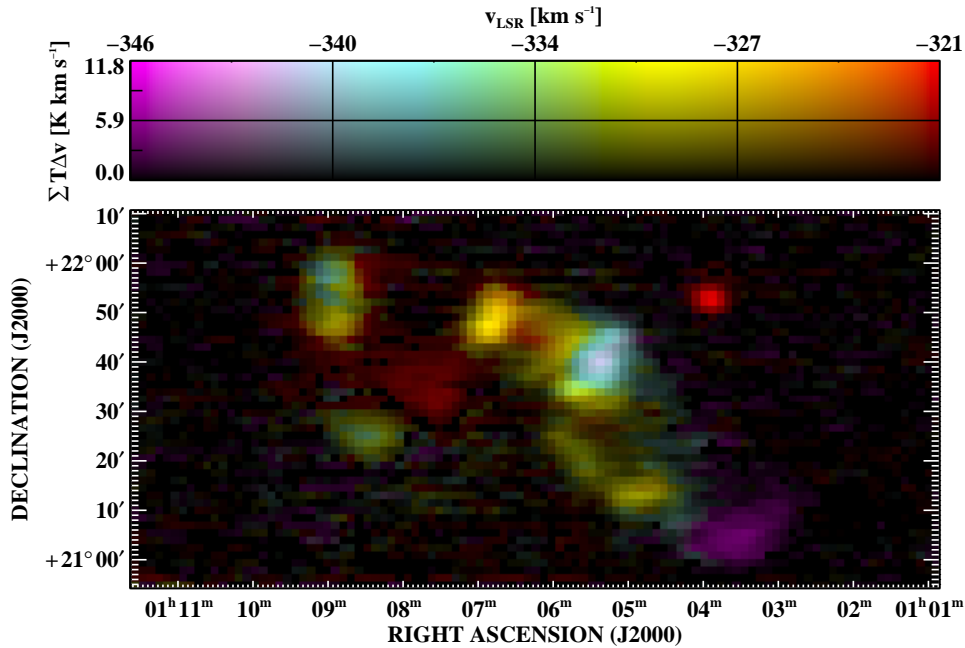}
\caption{High-sensitivity, wide-field H {\sc i} map of the region
containing the HVC and LGS~3.  Brightness represents integrated
intensity and the color scale represents velocity.  LGS~3 is the red
source in the upper right; the HVC is the double-lobed object above
the center of the map.  The faint features to the east and southwest
of the HVC that were barely visible in our original data now stand out
clearly.  Both are (mostly) spatially continous structures that
connect to the main body of the HVC, supporting our interpretation of
them as tidal tails that are being torn off of the HVC by LGS~3.
\label{lgs3nov02}}
\end{figure}

In August 2002, we used another Arecibo observing run to make a new,
more sensitive map of LGS~3 and the HVC.  This map, displayed in
Figure \ref{lgs3nov02}, confirms all of the features seen in our
original data.  The tidal tails are now visible in detail, and their
connection to the main body of the HVC is evident.  The tails are
somewhat more extended than could be seen in the first map, and
additional observations over a wider field have revealed that the
southwest tail continues on beyond the edge of the map shown in Figure
\ref{lgs3nov02}.  The new data are fully consistent with the scenario
proposed by Robishaw et al. (2002) of a tidal interaction between the
HVC and LGS~3.

\section{Conclusions}

Given these findings, what can we conclude about the nature of this
HVC?  We have presented several arguments that it is probably located
hundreds of kiloparsecs away, most likely at 700 kpc.  The HVC appears
to be undergoing a tidal interaction with LGS~3, which is stripping
away a substantial portion of its neutral gas.  If the HVC is in the
Local Group it is dark matter-dominated, regardless of its exact
distance.  And, it does not contain any stars (Robishaw et al. 2002).
This set of properties is exactly what is expected of the missing dark
matter satellites that are predicted by simulations.  We propose that
this HVC is the first observed representative of this population of
missing objects.

\acknowledgements{This research was supported by NSF grant
AST-9981308.}

\end{document}